\documentclass[a4paper]{article}

\usepackage{INTERSPEECH2019}

\title{Fine-grained Language Identification with Multilingual CapsNet Model}
\name{ Mudit Verma, Arun Balaji Buduru }

\address{Indraprastha Institute of Information Technology Delhi}
\email{mudit.verma2014@gmail.com, arunb@iiitd.ac.in}

\begin{document}

\maketitle
\begin{abstract}
Due to a drastic improvement in the quality of internet services worldwide, there is an explosion of multilingual content generation and consumption. This is especially prevalent in countries with large multilingual audience, who are increasingly consuming media outside their linguistic familiarity/preference. Hence, there is an increasing need for real-time and fine-grained content analysis services, including language identification, content transcription, and analysis. Accurate and fine-grained spoken language detection is an essential first step for all the subsequent content analysis algorithms. Current techniques in spoken language detection may lack on one of these fronts: accuracy, fine-grained detection, data requirements, manual effort in data collection \& pre-processing. Hence in this work, a real-time language detection approach to detect spoken language from 5 seconds' audio clips with an accuracy of 91.8\% is presented with exiguous data requirements and minimal pre-processing. Novel architectures for Capsule Networks is proposed which operates on spectrogram images of the provided audio snippets. We use previous approaches based on Recurrent Neural Networks and iVectors to present the results. Finally we show a ``Non-Class'' analysis to further stress on why CapsNet architecture works for LID task.
\end{abstract}

\section{Introduction}
\label{sec:intro}

\noindent  Language Identification (LID) systems are daily used in several applications such as intelligent personal assistants, emergency call routing or multilingual translation systems, where the response time of a fluent native operator might be critical. Currently, the users of these systems are required to select their preferred language for speech recognition. However, these systems should perform a sensible preprocessing step to infer the spoken language automatically using a Language Identification System. Traditional LID systems rely on domain experts in the field of audio signal processing for handcrafted feature selection and extraction from audio samples which requires tiring human effort. Deep neural networks have become one of the best-performing methods for a range of computer vision tasks, such as image classification \cite{chung2014empirical,simonyan2014very}, however, recently introduced Capsule Networks \cite{sabour2017dynamic}, shown to work well on image classification. Other methods involving Gated Recurrent Units with Attention \cite{luong2015effective,zhang2017gru,ba2014multiple} have been shown to work for time series data such as audio.


In this paper, we address the problem of language identification from a computer vision perspective to exploit the benefits of Convolutional Architectures. We extract the target language of a given audio sample by the CapsNet architecture. Our major contributions through this work are (1) A novel architecture for fine-grained language identification with Capsule Networks, (2) Reduced the dataset requirements for training purpose since the architecture is capable of handling languages with minimal dataset availability or collection, and  (3) Non-Class analysis, which essentially lets our architecture know when a given input does not belong to any of the languages for which our model is trained to detect. (4) Simple experiment to show that our Model is multilingual. 


\section{Related Work}
\label{sec:prior}

Over recent years, i-vector based framework \cite{7732177, Li2013SpeakerVU, Zeinali2016DeepNN} has been used for ASR and speaker verification in which each utterance is projected onto a total factor space and is represented by a low-dimensional feature vector. \cite{Dehak2011LanguageRV} takes this further to use dimensionality reduction over i-vectors followed by GMM \cite{burget2006discriminative} or SVM \cite{campbell2006support} classifiers for LID. \cite{Verma2015} explains the research work related to i-vectors and benefits of Mel-frequency cepstral coefficients (MFCC) in speech processing tasks. The type of input choice, in the image domain, is usually between spectrograms and Mel-frequency plots. A typical spectrogram uses a linear frequency scaling, so each frequency bin is spaced the equal number of Hertz apart. The Mel-frequency scale, on the other hand, is a quasi-logarithmic spacing roughly resembling the resolution of the human auditory system; this makes the MFCC features more biologically inspired. So essentially the major difference is whether one log the Mel-frequency spectrogram or not. Hence one of the major philosophical reasons, in addition to the empirical results we discuss later, to choose Spectrogram as our choice of input is to let the Neural Network learn the complex representations itself and not to impose them in any way. \cite{10.1371/journal.pone.0194770} tried LID through an ``extreme learning machine'' approach and used a small dataset with for 8 languages, suffers from poor recall and precision rates. \cite{li2005spoken} which obtains similar accuracy as ours but with triple the training dataset size and significant efforts in feature extraction. \cite{10.5120/ijca2016910924} suggests other feature extraction methods.

These models overlook efforts in data collection and processing, using complex feature representations, and assert faith in manual feature selection and domain expert knowledge. We solve the first concern by using raw spectrogram images, with no preprocessing, apart from the generation of spectrograms itself. Preprocessing is a crucial step which can result in significant improvements as shown by \cite{frederiksen2018effectiveness}. However, this requires considerable effort and time. Neural Networks are known to be Universal Function Approximators \cite{DBLP:journals/nn/HornikSW89}, hence, can handle much more complicated function mappings than other primitive approaches, which justifies their use in our solution. Lastly, use of Convolutional Neural Networks (CNN) as feature extractors is well known and studied especially in the field of Image Processing, so solving LID problem in image domains helps us to exploit this fact about CNN. 

\cite{gonzalez2014automatic,lopez2014automatic} applies feed forward neural network for the LID Task at the acoustic frame level. FDNN however, ignores the sequential nature of utterance sequences.
\cite{geng2016gating} proposes a variant of Gated Recurrent Units called Gating Recurrent Enhanced Memory Networks for temporal modeling of acoustic features composed of perceptual linear prediction coefficients and pitch coefficients. However, they have audio clips of 30 seconds as well in the training dataset which itself is large (NIST LRE 2007). \cite{garcia2016stacked} also deals the problem of LID task using i-vectors and Bottleneck features \cite{fer2015multilingual} which requires preprocessing of data like vocal tract length normalization, RASTA filtering, conversion to MFCCs shifted delta cepstra (SDC), etc.

\cite{lozano2015end} perform Language Identification with the help of CNN's by transforming data to MFCC-SDC features. Building on that, \cite{bartz2017language}, uses various CNN \& LSTM based approaches which involves VGG Network \cite{simonyan2014very}, inception \cite{Szegedy2016RethinkingTI}, residual Networks \cite{7780459}, however large dataset size (more than 1500 hours) is used for training the models, hence the architecture cannot be used for regional languages for which large amounts of data is unavailable, or which requires much human effort for data collection. Retaining spatial relative organization of image pixels ahead in the layers becomes crucial for the task of classifying languages, since there might be only subtle differences in the choice of words, something other than MaxPooling (used in Convolution Blocks) is required, which can very will see the bigger picture and takes the organization of image components into account for classification. This served the motivation for Capsule Networks \cite{sabour2017dynamic}. We further classify for Non-Class or Out-Of-Set languages(OOS). \cite{behravan2016out} implements this through Kolmogorov- Smirnov (KS) test, however, our motive is to test how well the output vectors/probabilities learn the language to discard other languages to Non-Class set via a thresholding mechanism.

\section{Dataset}
\label{sec:dataset}
\frenchspacing

As noted by \cite{bartz2017language}, there are no freely available large scale datasets for LID task with multiple languages, those such as NIST LRE \footnote{https://www.nist.gov/itl/iad/mig/language-recognition}, OGI Multilanguage Corpus \footnote{https://catalog.ldc.upenn.edu/LDC94S17} are available behind a paywall. We, therefore, create our dataset for training and testing purposes as described below. However, we could find openly available datasets for English ( OpenSLR LibreSpeech ASR Corpus trainset of clean 100 hours \footnote{http://www.openslr.org/12/} based on LibriVox's audio books.) and Arabic\footnote{http://en.arabicspeechcorpus.com/}. For both of these datasets, we randomly sample our testing set of 5 hours each. 

As mentioned, an audio dataset has been created for 10 languages in two sets. Classification Set: \{Arabic, Bengali, Chinese Mandarin, English, Hindi\}. 70 hours of training and 30 hours test for each language. Non-Class Set: \{Turkish, Spanish, Japanese, Punjabi, Portuguese\} 30 hours for Non-Class Analysis for each language. Dataset involves audio recordings of various local and global news interviews, speeches, discussions etc. of lengths varying from 20 seconds to 2 hours. The Dataset shows the below important characteristics in favor of creating a good LID system.

\noindent $\bullet$ Ease in availability and collection. All data has been downloaded from YouTube such as interviews, speech, etc. videos. \\
$\bullet$ Use of raw data favors an extension to regional languages for which obtaining data is a difficult task. \\
$\bullet$ Natural Noise. Firstly, Heard/Not-Understandable: Noise of the crowd, cheers, slogans, chants, etc. which are spoken but not clearly understood. Secondly, Heard/Understandable: Multiple languages are spoken in the clip. Lastly, Unheard: Chimes, laughter, instrumental songs, etc. which does not involve any spoken language.

For dataset images, We use lossless WAVE format, as this allows manipulations without deterioration in signal quality. We convert audio snippets to spectrogram representations. The audio is first converted to an image and then clipped to 5s/10s snippets, therefore creating two datasets, one with 5s clips and the other with 10s clips. Spectrograms are discretized using a Hann \cite{6768513} window and 129 frequency bins (or 64 bins) along the frequency axis. The time axis is rendered at two rates, 5 pps (pixels per second) for 5-second clips and 50 pps for 10-second clips using SoX \footnote{http://sox.sourceforge.net/Main/HomePage}. This would generate images of sizes (64,50) and (500,129) respectively. The resulting images are then resized to shape (32,25) using OpenCV \cite{bradski2000opencv}.


\section{Proposed Architecture}
\label{sec:algo}

\begin{figure}[t!]
\centering
\includegraphics[scale=0.24]{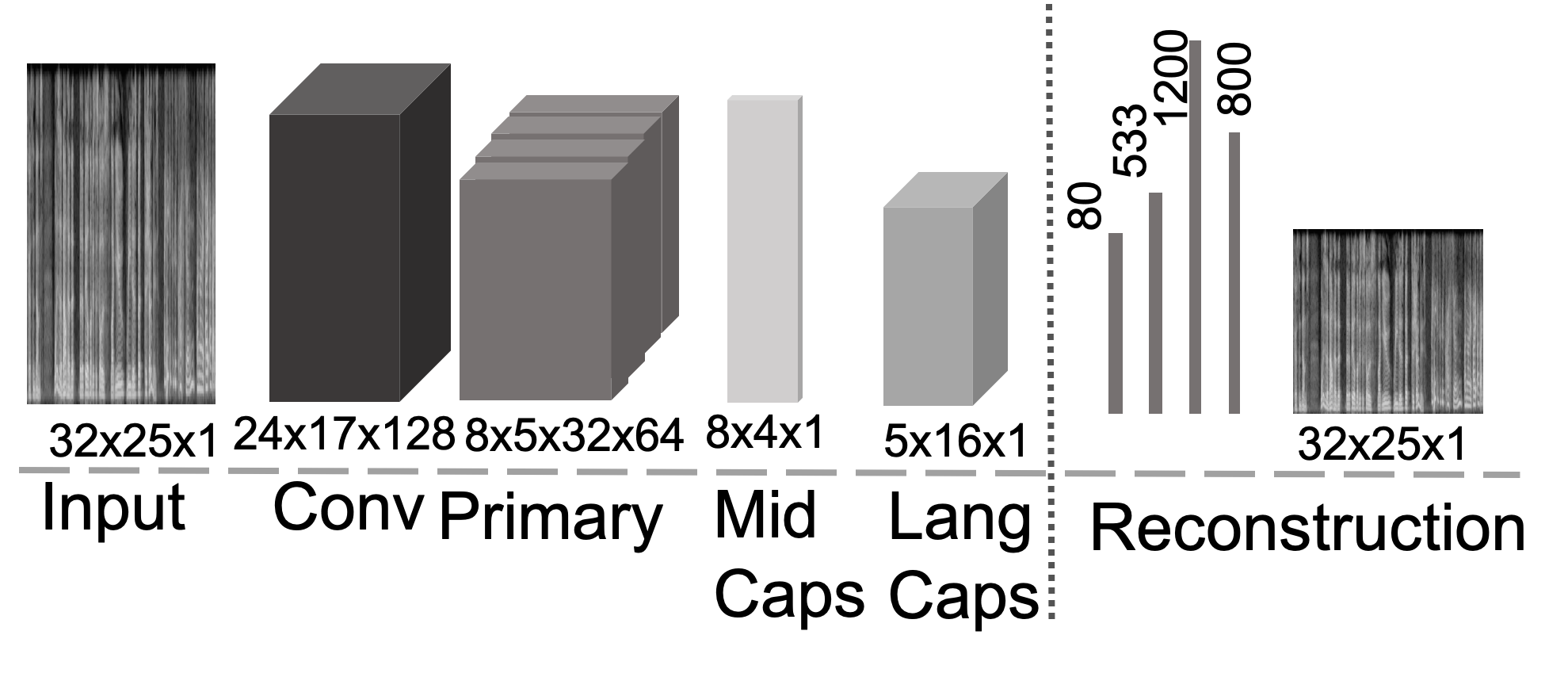}
\caption{Proposed CapsNet architecture for LID.}
\label{fig:capsnetArch}
\end{figure}

We propose a deep learning architecture, figure \ref{fig:capsnetArch} based on Capsule Networks (CapsNet), explain its components followed by Non-Class Analysis which reveal further insight about the learned neuron values.



Our variant of CapsNet has 2 parts: encoder and decoder as in Figure \ref{fig:capsnetArch}. The first 4 layers constitute the Encoder, and the final 3 makes up the Decoder. The Encoder takes as input a 32x25 Spectrogram and learns to encode it into a 16-dimensional vector of instantiation parameters; this is where the capsules perform their job. The output of the network during prediction is a 5-dimensional vector of lengths of LangCaps' outputs. The input image goes through the encoder phase until LangCaps Layer with Dynamic Routing Applied in the MidCaps and LangCaps Layer. Classification is done via magnitude computation of LangCaps Vectors.

\subsection{Encoder}

A convolution layer is used to detect basic features in a 2D image. In the CapsNet, the convolutional layer has 128 kernels with a size of (9,9,1) and stride 1, followed by ReLU. Primary Caps Layer has 64 primary capsules, and it takes basic features detected by the convolutional layer and produces combinations of the features. The layer has 64 ``primary capsules'' that are very similar to a convolutional layer in their nature. Each capsule applies 32, (9,9,128) convolutional kernels, with stride 2, to the input volume and therefore produces (8,5,32) output tensor. This is input to the MidCaps layer which uses routing and has been introduced to bring the concept depth in capsule layers and capture non-linearities. Then finally the LangCaps Layer is used for language classification. This layer has 5 capsules, one for each language. Each capsule takes as input an (8,4,1) tensor. 



\subsection{Loss Function, Decoder and Regularization}

\begin{equation}
\label{lossfun}
    L_c = max(0, m^+ -  \left\lVert v_c \right\rVert)^2 + \lambda(1-T_c)max(0,\left\lVert v_c \right\rVert - m^-)^2
\end{equation}

The complete Loss function is a sum of Margin Loss (equation \ref{lossfun}) and Reconstruction Loss. During training, one loss value will be calculated for each of the 5 vectors for each example and then added together to calculate the final loss. Decoder takes as input a 16-dimensional vector from the correct LangCap and learns to decode it into a spectrogram. It first expands the output of the encoder and then contracts it back to the required shape of (32,25) through three fully connected layers, which is used for calculation of Reconstruction loss: the regularization term. Such an expansion lets the network view the data in other higher order dimension before contracting it back.

The equation \ref{lossfun} works as follows. For a correct language label, say 1, means that the first capsule in LangCap is responsible for encoding the presence of the language 1. For this LangCap's loss function $T_c$ will be 1 and for all remaining four LangCaps $T_c$ will be 0. When $T_c$ is 1 then the first term of the loss function is calculated and the second becomes zero. For our model, in order to calculate the first LangCap’s loss, we take the norm of the output vector of this LangCap and subtract it from $m^+$, fixed at 0.9. The resulting value is kept only when it is greater than zero, otherwise, 0 is returned. In other words, the loss is zero only when the correct LangCap capsule is of maximum value with probability less than 0.9.

\subsection{Non-Class Detection}

Non-Class is an important part of this LID system showing what languages are understood by the classifier. It shows us how ``good'' are the vector values/probabilities given by the architectures.
As mentioned in \cite{sabour2017dynamic}, the individual capsule values signify space variations or specific digit variations in the MNIST handwritten digit dataset, like the stroke thickness. Building an analogy, we expect the capsules for our CapsNet architecture to hold some meaning. This motivates us to perform non-class analysis, to check the ability of our architecture to detect out-of-set languages by using the capsule neuron values.

To do this, we sample out 20 hours of data from the training set for each language, and obtain norm of capsule neurons (A capsule is itself a set of neurons), at the last layer, before application of softmax activation for classification. For all the true positive results, we then find the minimum value of the capsule vector norm, for each language, and use these as our thresholds. At the testing time, given a language sample, we obtain the network classification as well as the capsule vector norm (as mentioned above), and compare it with the thresholds for each language. If the vector norm is less than the thresholds, we state the sample to belong to an out-of-language set. 


\section{Baselines}
We compare the performance of our architecture with three types of models.
First, our Baseline-1 is CRNN model from \cite{bartz2017language} for comparison in image domain. Second, Baseline-2 is iVector + MMI + SVM-GSV model \cite{Dehak2011LanguageRV} run on 5 second clips.
The third is our CGA architecture inspired by the works of \cite{bartz2017language} as explained below.

\begin{figure}[t!]
\centering
\includegraphics[scale=0.24]{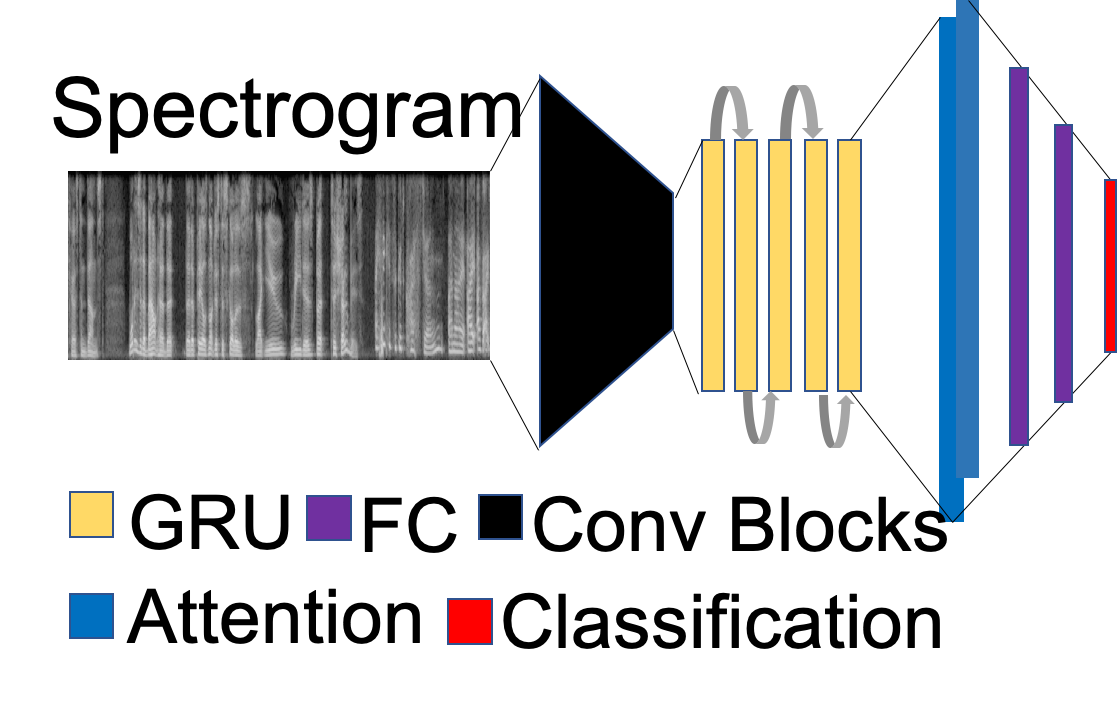}
\caption{\label{fig:cgaArch}CNN+bi-GRU+Attention (CGA) architecture for LID}
\end{figure}

\subsection{CNN+bi-GRU+Attention (CGA)}
Figure \ref{fig:cgaArch} shows CGA architecture. This is a variant of CRNN as in \cite{bartz2017language}. CGA is composed of Convolutional Neural Blocks (CNN and MaxPool) with Bi-directional Gated Recurrent Units and added Attention before classification layers. Five convolutional blocks are used for feature extraction, first three blocks with stride 1 in Convolutional Layer and stride (2,2) in MaxPool Layer. The fourth block has stride (2,1) at MaxPool layer, and finally, for the last block, stride for the MaxPool layer is (2,2). Number of convolutional filters used are of size [16,32,64,128,256] respectively for each block. We then add Bi-Directional Gated Recurrent Units Layer with 512 outputs in either direction and pass these outputs to the Bi-directional GRU. Then our Attention block takes input from the Bi-Directional GRU of the shape (None,53,1024). We Permute the second and first dimension and add a fully connected layer with softmax. Permuting the shape back to (None,53,1024) gives us the attention vector, and the values are attention probabilities. Finally, the attention probabilities are used and, combined with fully connected layers at the end; where classification is performed.

\section{Results}
\label{resultsSection}


 All the models have been trained and tested on our dataset as in Section \ref{sec:dataset}. We also showcase results on CGA-1: CGA with Bi-LSTM \& no attention and CapsNet-1: CapsNet without MidCaps Layer.

\begin{figure}[t!]
\centering
\includegraphics[scale=0.19]{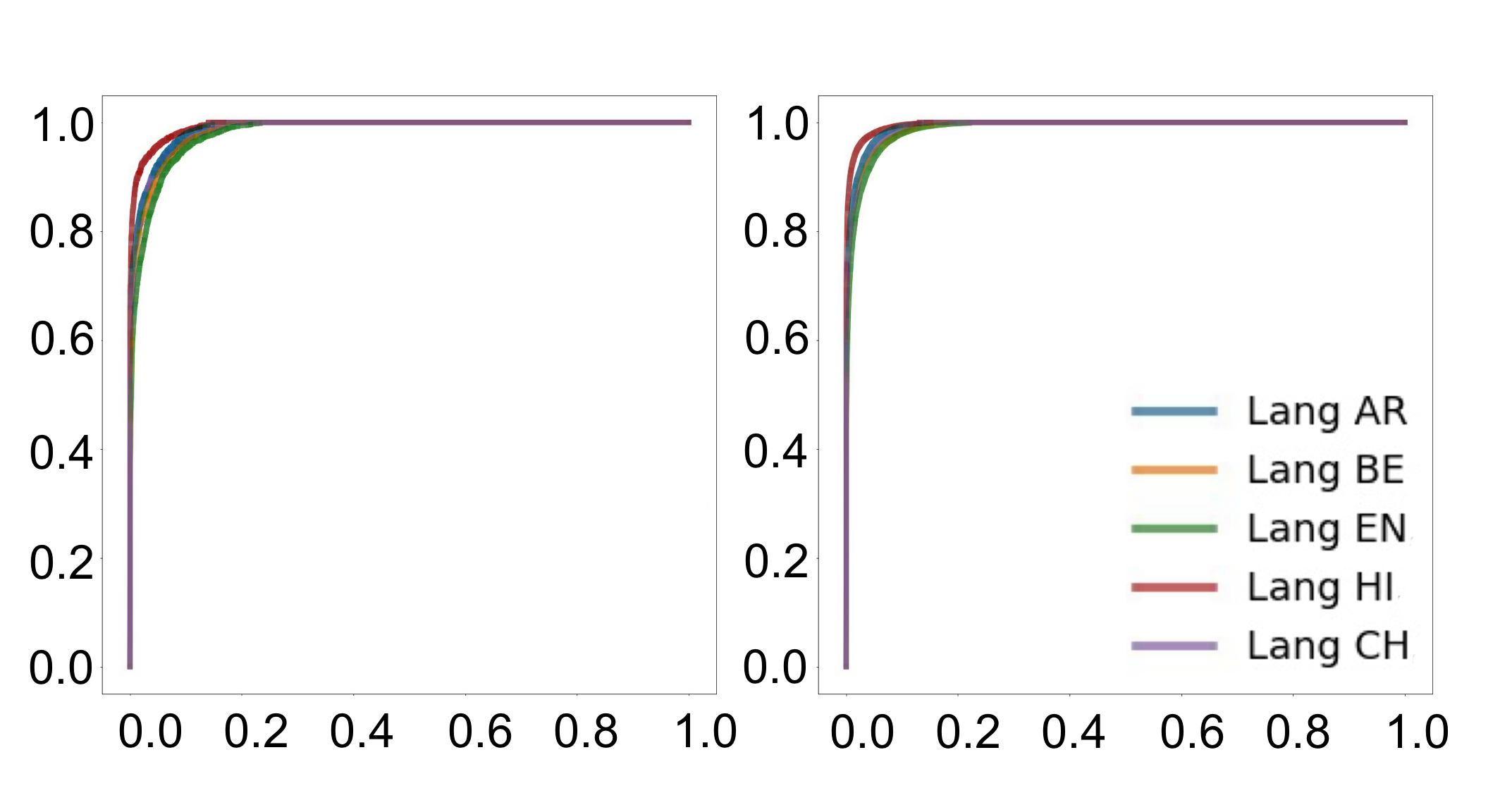}
\caption{\label{fig:capsROC}ROC Curves for training (left) and testing data (right) of Classification set languages for CapsNet.}
\end{figure}

Figure \ref{fig:capsROC} clearly shows the smooth ROC curves for CapsNet giving AUC=(0.98-0.99) for both training and testing sets indicating no overfitting. Evident from Tables \ref{tab:accuracyTable} \& \ref{tab:Precision}, this architecture has least overfit the data, and given the diversity in the language phonetics it has very well generalized for only 70 hours training data for each language. With high Precision \& Recall Rates, this has proved to be better than all other at half the input clip size. CapsNet surpasses final accuracies of CGA, CGA-1 in just over 3 hours of training time. CapsNet-1 works as well as the CGA, but it does not overfit, validating our confidence in using Capsule Networks. CapsNet architectures were also tested on Dataset with Mel-frequency generated images for the same audio data set and the average testing accuracies achieved were a little over than 70\% proving our hypothesis of using Spectrograms for this problem. Among the CNN+RNN family, CGA reached the maximum Test Accuracy. Hence, using Attention helps the most to boost the overall Accuracy. Moreover, although the issue of model overfitting, sourcing from lack of dataset, is still prevalent for all RNN based models, CGA gives a smoother training plot. iVector based baseline-2 performs slightly worse than baseline-1 but is not over-fitted, however, CGA, CapsNet and all their variants beat both the baselines in accuracy and AUC percentage.

\begin{table}[h]
\caption{\label{tab:accuracyTable}Comparison of different Architectures. 
}
\centering
\resizebox{0.9\linewidth}{!}{%
\begin{tabular}{|l|l|l|l|l|}
\hline
\textbf{\begin{tabular}[c]{@{}l@{}}Architecture (Clip Seconds)\end{tabular}} & \textbf{\begin{tabular}[c]{@{}l@{}}Acc. Test\end{tabular}} & \textbf{\begin{tabular}[c]{@{}l@{}}Acc. Train\end{tabular}} & \textbf{\begin{tabular}[c]{@{}l@{}}Non-Class\end{tabular}} & \textbf{\begin{tabular}[c]{@{}l@{}}AUC Test\end{tabular}} \\ \hline
CGA-1      (10)             & 63.4 & 89.4 & 56.22 & - \\ \hline
\textbf{CGA         (10)}   & \textbf{71.04} & \textbf{95.12} & \textbf{68.22} & \textbf{88.2} \\ \hline
CapsNet-1 (10)              & 74.58 & 77.02 & 63.16 & - \\ \hline
\textbf{CapsNet    (05)}    & \textbf{88.76} & \textbf{91.8} & \textbf{73.04} & \textbf{98.2} \\ \hline
Baseline-1      (05)             & 62.98 & 85.35 & - & 79.8 \\ \hline
Baseline-2 (05)             & 57.24 & 60.48 & - & 74.1 \\ \hline
\end{tabular}}
\end{table}

The CapsNet model, trained on our dataset when evaluated on 5 hours of the freely available dataset (converted to 5-second clips) for English and Arabic, as explained in section \ref{sec:dataset}, we achieved accuracy values of 85.24\% as 83.61\%. Since a little dip in accuracy values is expected due to different training and testing set distributions, the model gives a respectable accuracy value in comparison with 88.76\% as mentioned in table \ref{tab:accuracyTable}.

\begin{table}[h]
\caption{\label{tab:Precision} Comparison Measures between CapsNet and CGA. Each cell is (Test/Train). }

\centering
\resizebox{0.9\linewidth}{!}{%
\begin{tabular}{|c|c|c|c|c|c|c|}
\hline
\textbf{Lang} & \multicolumn{2}{c|}{\textbf{Precision x 10\textasciicircum{}-2}} & \multicolumn{2}{c|}{\textbf{Recall x 10\textasciicircum{}-2}} & \multicolumn{2}{c|}{\textbf{F1-Score x 10\textasciicircum{}-2}} \\ \cline{2-7} 
 & \textbf{Caps} & \textbf{CGA} & \textbf{Caps} & \textbf{CGA} & \textbf{Caps} & \textbf{CGA} \\ \hline
\textbf{AR} & 89/92 & 75/94 & 89/93 & 75/95 & 89/92 & 75/95 \\ \hline
\textbf{BE} & 87/91 & 69/94 & 90/93 & 72/96 & 89/92 & 71/95 \\ \hline
\textbf{EN} & 88/93 & 62/96 & 82/86 & 53/91 & 85/89 & 57/93 \\ \hline
\textbf{HI} & 92/94 & 78/96 & 93/95 & 81/97 & 92/94 & 80/97 \\ \hline
\textbf{CH} & 87/90 & 69/94 & 90/93 & 72/95 & 89/92 & 80/94 \\ \hline
\textbf{Avg} & \textbf{89/92} & \textbf{71/95} & \textbf{89/92} & \textbf{71/95} & \textbf{89/92} & \textbf{71/95} \\ \hline
\end{tabular}}

\end{table}

\begin{figure}[h]
\centering
\includegraphics[scale=0.19]{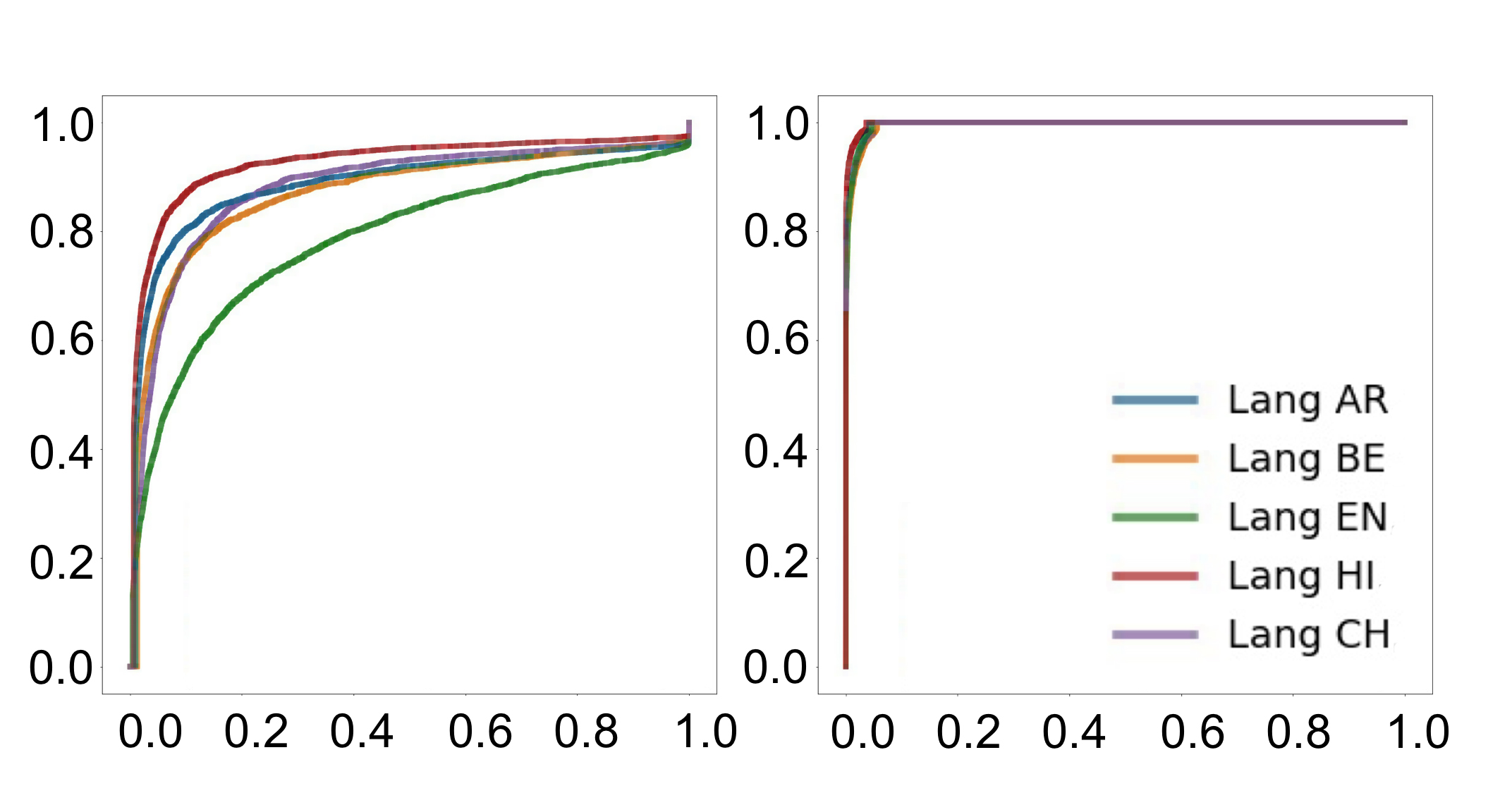}
\caption{\label{fig:gruROC}ROC Curves for training (left) and testing data (right) of Classification set languages for CGA.}
\end{figure}

Evident from Figure \ref{fig:gruROC}, AUC is poor for the Test set, and is extremely high on Train set for CGA. AUC is very high when either the model is a good classifier, or it has over-fitted the data, here, however, it has over-fitted. Table \ref{tab:Precision} makes it clear that significantly lower test rates for Precision, Recall, and F1-score than test values, that GRUs require more training data and also highlights that these class of models are sensitive to noise in comparison with CapsNet. On the other hand, the relative values of Precision, Recall \& F1-Scores are similar for Test \& Train for CapsNet, a mark of a well fit model. Scores are highest for Hindi and least for English suggesting that deep learning models suffer when diverse languages are present. Hence a single classifier to identify all languages (or a large set of distinct languages) may have poor accuracy. 

\begin{figure}[h]
\centering
\includegraphics[scale=0.30]{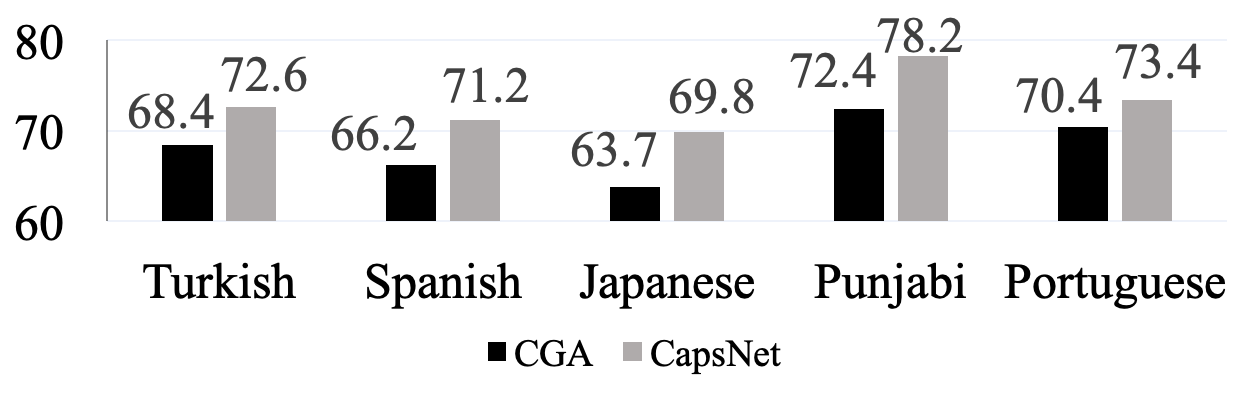}
\caption{\label{fig:nonClass}Non-Class Detection on Non-Class language set }
\end{figure}


Figure \ref{fig:nonClass} percentages correspond to how accurately Non-Class language set was classified as from an unknown class. From Figure \ref{fig:nonClass} and Table \ref{tab:accuracyTable}, we note that CapsNet performs best for Non-Class detection, where best detection occurred for Punjabi followed by Portuguese, and worst for Japanese wherein only 69.8\% of the samples were detected as Non-Class. We perform Non-Class analysis on our proposed models and their variants since the motive to involve Non-Class, also, to classifying out-of-set languages, is to evaluate how well a model understands the input languages and their fine-grained distinctions, such that their output values directly showcase the difference. Our CapsNet variant performs better due to a finer explanation of the presence of language by the classification neurons before softmax.

\begin{figure}[htb]
\begin{minipage}[b]{0.49\linewidth}
  \centering
  \centerline{\includegraphics[width=2.1cm]{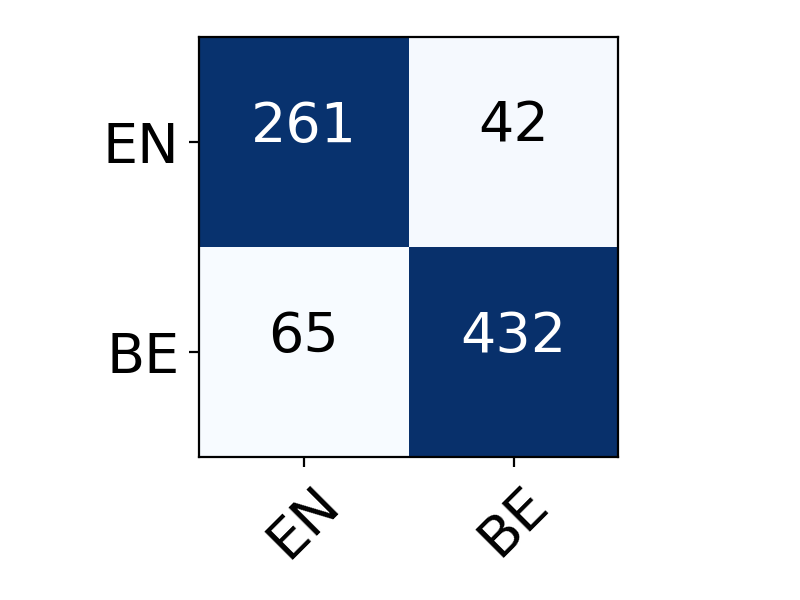}}
  \centerline{(a) EN/BE Multi Language}\medskip
\end{minipage}
\hfill
\begin{minipage}[b]{0.49\linewidth}
  \centering
  \centerline{\includegraphics[width=2.1cm]{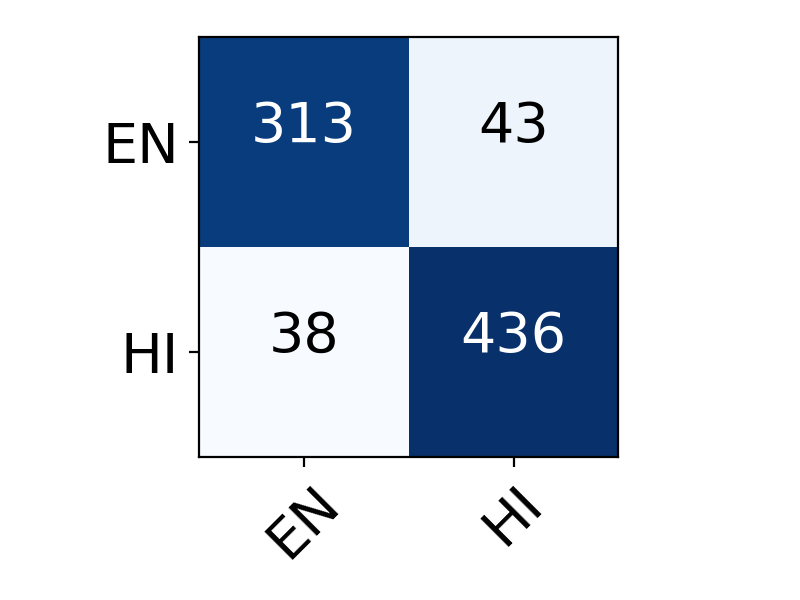}}
  \centerline{(b) EN/HI Multi Language}\medskip
\end{minipage}
\caption{Confusion Matrices}
\label{fig:res}
\end{figure}

Our CapsNet is multilingual \cite{toshniwal2018multilingual}. To show this, we performed a simple experiment. We manually sampled 16 audio clips of 250 seconds each which consists of 2 languages (EN/HI) or (EN/BE). We split each of the clip into 5 second snippets and manually label each snippet. Testing the CapsNet on these snippets, we found the following. For EN/HI, (800 total snippets, 356 EN, 474 HI) and for EN/BE (800 total snippets, 303 EN, 497 BE). From Fig. \ref{fig:res} we see that the results are similar to accuracy values achieved before.

\section{Conclusion \& Future Work}
In this work, we propose a novel capsule based architecture, CapsNet for the LID task through image domain. The architecture is shown to work with exiguous data requirements, with little overfitting. Further, a non-class analysis shows that the norm values of capsules better represent languages. We also show results of the model on certain openly available datasets for English and Arabic, trained on our dataset.
Further improving the CapsNet Model through the use of RNN cells with the Capsules involved. Possibly the approach of using GRUs with Attention can be applied to capsules as well. Moreover, advanced unsupervised/supervised methods may be used for Non-Class analysis.


\vfill\pagebreak

\bibliographystyle{IEEEtran}

\bibliography{mybib}


\end{document}